# New Uses for the Kepler Telescope:
# A Survey of the Ecliptic Plane For Transiting Planets and Star Formation.
## A White Paper submitted by

Charles Beichman, David Ciardi, Rachel Akeson, Peter Plavchan (NExScI),
Steve Howell, Jesse Christiansen (NASA/Ames),
Stephen Kane (San Francisco State University),
Ann Marie Cody, John Stauffer (IPAC),
Gautam Vasisht (JPL), Kevin Covey (Lowell Observatory)

3 September 2013

## I. Introduction

The Kepler mission has revolutionized the field of exoplanets with its 4 years of observing a single 100 sq. deg. field in the constellation Cygnus. By observing 150,000 stars continuously with <0.03 mmag precision, Kepler was able to find over 130 confirmed planets and 3,500 candidates. These planets range in size from less than an Earth-radius to twice the radius of Jupiter. The legacy of Kepler will last for decades to come, providing statistical information on entire populations and individual targets for detailed study.

With the loss of two reaction wheels, the period of Kepler's ultra-high precision photometric performance is at an end. Yet Kepler retains unique capabilities impossible to replicate from the ground or with existing or future space missions. This White Paper calls for the use of Kepler to conduct a survey in the ecliptic plane to search for planet transits around stars at high galactic latitudes and to study star forming regions to investigate physics of very young stars not studied by Kepler in its prime mission. Recent analysis by the Kepler project indicates that the spacecraft's best pointing will be possible in the ecliptic plane.

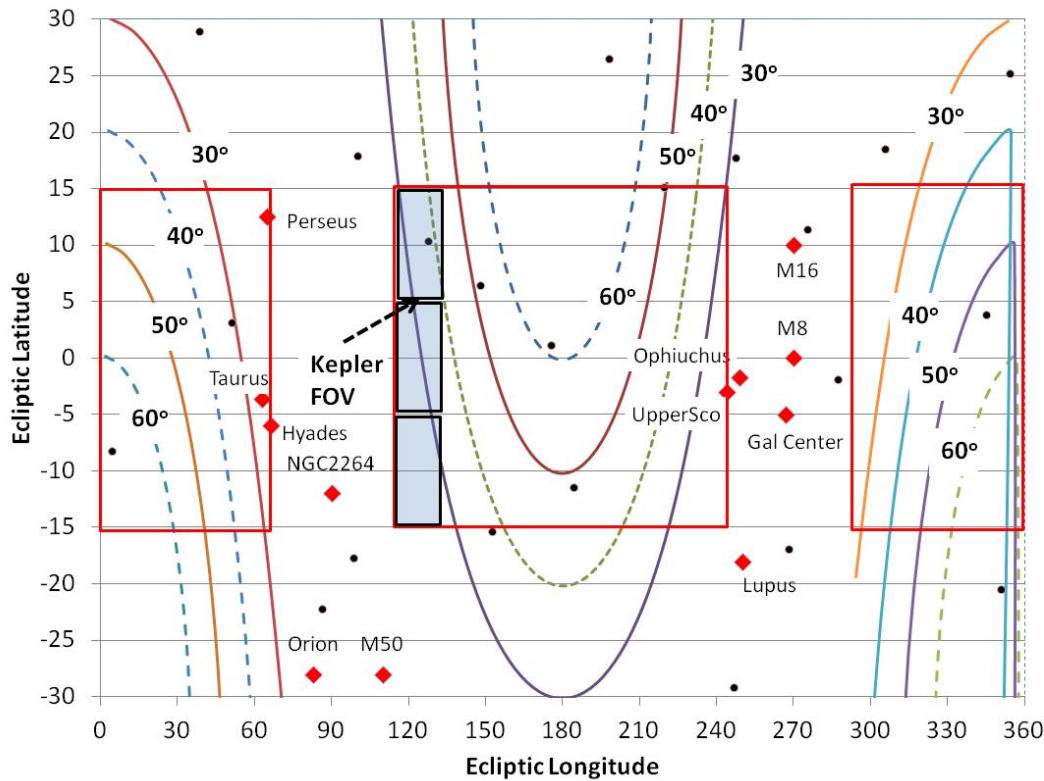

Figure 1. Kepler could survey bright FGK stars and M stars in up to 40 high galactic latitude fields (|b|>30 deg) in the ecliptic plane in a 3 yr extended mission. The red boxes denote high latitude areas Kepler could survey in ~3 years with illustrative Kepler fields shaded in blue. Solid and dashed lines denote parallels of galactic latitude. When high galactic fields are not available, Kepler could study star forming regions or other galactic plane objects (red diamonds). Known RV planetary systems with a high probability of transiting are show as black circles.

Numerous other scientific interests might be addressed with the community providing ancillary targets during the high galactic latitude transit survey, or alternatively, selecting primary targets in which case transit searches could be being carried out looking at whatever bright stars happen to be in the selected field(s).

In its extended mission, Kepler would be used in an operational manner very similar to its present mode of operation (long term staring at a few fields per year) to send down pixel postage stamps around targets of interest. These data can be processed through the initial stages of the existing Kepler pipeline with little or no modification before being archived for use by investigators to achieve their scientific goals.

We propose a 3 year program to survey between 5 and 10% of the sky located near the ecliptic plane (where pointing drifts are minimal) and at high galactic latitudes, |b|> 45 degree (where confusion effects are minimized). The proposed survey could examine roughly 20 fields (ecliptic latitude<±5 deg) for 60 days each or 40 fields (ecliptic latitude<±15 deg) for 30 days each. **The survey could yield hundreds of gas and ice giants orbiting host FGK stars and hundreds of ice giants and rocky (radius < 2 $R_\oplus$) Super Earths orbiting mid- to late M stars.** The trade between stare time vs. the number of fields would be made as part of a more complete proposal. In general, observing more fields yields a greater number of planets orbiting brighter stars, while a longer duration stare per field yields longer orbital periods, in particular probing more fully habitable zone orbits around M stars.

As discussed below, the average planet host star in the extended mission would be ~10 times brighter (2.7 mag) than those in the prime mission, greatly easing the challenge of validation and

| Table 1. Comparison of Various Facilities | | | | | | | |
|---|---|---|---|---|---|---|---|
| Facility | Year of Operation | Estimated Limiting Photo-metric Precision (mmag) | Magnitude to achieve limiting precision in 15 minutes (mag) | Instan-taneous FOV (sq. deg.) | # Stars at high cadence | Min/Max Dwell Time per year (days) | Observing Completeness |
| *Kepler (2 wheel)* | *2014* | *1.0 0.5* | *R=17 16* | *100* | *>20,000 per field (15 minutes)* | *Continuous For 30-60d* | *96% (lose 4 hours in 4 days)* |
| Warm Spitzer | 2011, ongoing | 1 | [4.5]=14 | 0.007 | 5 /hour | Continuous for 45d | 20% (repeat 5 stars per hour) |
| TESS | 2018-2020 | 0.06 | I<7 | 4x23^2 ~2100 | >250,000 | 27d in ecliptic plane | >90% (lose 1 day in 14 days) |
| CHEOPS | 2017-TBD | 0.05 | V<9 | pointed | ~100/yr | Continuous for 1~30 day | >90% |
| MEarth-N/S (ground) | 2010, ongoing | 1-3 | R<14 | pointed | 1,000 (4,000 total/yr) | 0.25day/day for 90d | <0.15 (day/night, weather, cadence) |
| HATNET-S (ground) | 2010, ongoing | 1-3 | R<14 | 128 sq. deg. x 3 sites | >500,000 | Up to 1day/day for 90d | <0.3 (weather) |

characterization and improving the suitability of new planets for spectroscopy by JWST and other facilities. In the case of the M stars, the larger aperture of Kepler compared to that of TESS will improve the yield of small planets orbiting those stars for which photon-limited signal to noise, not high photometric precision, is the limiting factor. Finally, in comparison with TESS, a Kepler survey of the ecliptic plane will fill in coverage gaps and extend the time baseline of observations of TESS's observations. The extended time baseline enables the confirmation of exoplanet candidates in tightly packed orbits through transit timing variations, where dynamical interactions can evolve over several years.

The conduct of the proposed survey would be highly uniform and repetitive for improved data quality, minimal operational cost, planned re-use of first few steps in Kepler pipeline software, and ease of data reduction. The ecliptic and galactic latitude geometric constraints for the transit survey can be met for roughly 2/3 of a year. In the remaining time, Kepler could stare at a set of community-selected fields (Figure 1), chosen to highlight specific scientific goals, such as potential transiting planets selected from short period Radial Velocity systems, studies of variability in star forming regions such as Taurus, Upper Scorpius, Ophiuchus, or Orion, or microlensing targets in the Galactic Bulge (see the White Paper by Gould and Horne). In addition to the targets selected for specific study, the community would be encouraged to suggest hundreds or thousands of additional targets for otherwise uncommitted "postage stamps" for monitoring on long or short cadences to investigate stellar variability or rotation, or extragalactic objects. Targets for these projects could be incorporated into the main observing program via a peer-reviewed process just as they were during Kepler's prime mission.

## II. Assumed Performance in 2 Wheel Mission

Even with the degraded performance expected with 2 reaction wheels, Kepler remains a powerful observing facility. Table 1 compares its predicted performance with other facilities, including Spitzer, TESS, CHEOPS, and two ground-based transit survey facilities, HATNET and MEarth.

- Compared to TESS's 10 cm telescope, Kepler's 1 m telescope offers a significant advantage in reaching a photometric precision of 1 mmag, e.g. R of 17 mag vs. 13 mag in 15 minutes (Jenkins et al 2010; estimated from Ricker 2013). This advantage is important for observing faint targets such as M stars or, more generally, objects in star forming regions or active galaxies. More important than just its sensitivity advantage, Kepler can point at individual fields with near 100% complete cadence for up to 60 days at a time to enable long term monitoring of tens of thousands of targets looking for transits, e.g. multiple transits of habitable zone planets orbiting M stars, rotational periods of main sequence stars, pre-main sequence binaries, young variable stars, or outbursting pre-main sequence objects. Finally, Kepler has the advantage of returning its results starting only a few months from now (compared with a 3-5 year wait for TESS) in time to produce

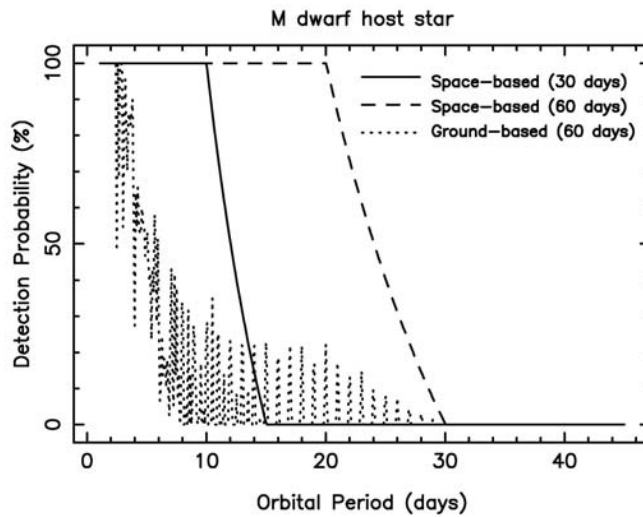

Figure 2. The probability of detecting 3 transits in a given observing duration as a function of orbital period. The solid line and dashed lines are for 30 and 60 day stare periods with Kepler. The dotted line simulates the probability of detection with a ground based survey with a 60 day observing period (8 hours per day with 50% weather loss). The curves reflect only visibility considerations, i.e. perfect signal to noise in both cases.

well vetted targets suitable for study by JWST and other facilities.

- Compared with Spitzer, Kepler's advantage in field of view (100 sq. deg vs. 0.007 sq. deg) results in an unsurpassable capability for the study of large numbers of variable sources spread over large regions of sky, simultaneously, when near continuous coverage is required. Spitzer's <0.03 mmag precision remains critical for the study of individual objects, however.
- Compared with ground-based transit surveys, Kepler offers long term, continuous coverage impossible to achieve when weather and diurnal effects are taken into account. Kepler's operation from space with a 1 m telescope will offer dramatic improvements in limiting magnitude and, we anticipate, in limiting precision compared with any ground-based wide angle survey. The absence of seeing and other atmospheric effects ("red noise") should permit Kepler, even in its 2 wheel mode, to achieve 1 mmag or better precision. The completeness for transits of various periods for Kepler and a typical ground-based survey is shown in Figure 2 and demonstrates the deleterious effects of interrupted sequences of measurements from the ground compared with Kepler's near-continuous data.

In what follows, we assume a limiting single measurement accuracy of 1 mmag and that this value is a random variable which will continue to average down with multiple observations. This is equivalent to assuming that Kepler observations will have a 1/f, or "Red Noise", floor that is <1 mmag. Thus, if Kepler achieves a noise value of 1 mmag on a bright star in 15 minutes (R<16 mag), then the limiting noise after 1 hour is 0.5 mmag. Because we expect the photometric errors to be correlated during a single pointing using a small number of pixels, we impose a noise floor of 0.3 mmag for the duration of a 4 day measurement period between momentum dump maneuvers. Measurements from one 4-day period to the next will use a different set of pixels due to the 20-30" blind pointing accuracy of the star trackers and should thus be uncorrelated. These assumptions are directly relevant to the transit investigations discussed below, but must of course be validated through on-orbit performance.

We further note that the goals of this white paper can be accomplished using tens of thousands of targets compared with Kepler's original 150,000 targets with significant advantages for mission planning and accommodating expected pointing performance (§IV.A).

## III. Science Opportunities

### A. Exoplanet Research

**Conduct a High Galactic Latitude Transit Survey.** While Kepler in its prime mission has greatly expanded the number of known transiting systems, there are gaps in its parameter space of stellar brightness and spectral type. By using Kepler to cover more than 40 times as much sky as in its prime mission, we can address these lacunae with expanded searches for planets orbiting M stars whose smaller radii and luminosity can yield planets of modest size in stellar habitable zones and FGK stars 10 times brighter than typical Kepler systems (Table 2). For these new surveys, we would concentrate on high latitude fields (|b|>45 deg) to reduce contamination from background stars by a factor of 5-10 which will greatly reduce the challenge of confirmation. To ensure the lowest level of pointing drift we would concentrate our observations to lie close the ecliptic plane. An ecliptic plane survey would complement the TESS survey which will have relatively poor dwell times in this part of sky, and possibly even gaps in its sky coverage during its prime mission.

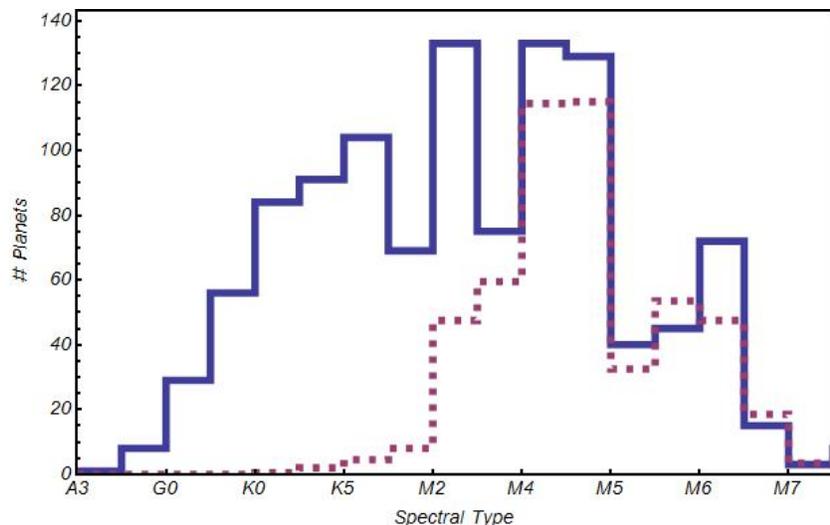

Figure 3. The yield of planets from a bright star Kepler survey in the ecliptic plane (340,000 stars in 40 fields observed 30 days each) is shown as a function of host star spectral type. Gas or icy giants are predominantly found around FGK stars (solid curve) while smaller icy or rocky planets (dashed curve) are seen predominantly orbiting M stars.

Assuming that *nearby* M stars are isotropically distributed, then the number of objects found in a survey will be $N \propto \Omega\ 10^{-0.6mag}$. The magnitude limit, *mag*, needed to reach N stars in a broad survey, $\Omega_1$, is brighter than the mag limit for a small survey, $\Omega_2$, by an amount $\Delta mag = 1/0.6\ log(\Omega_1/\Omega_2)$. By surveying 40 (20) fields (30 or 60 days each) in the extended mission compared to Kepler's single field in its prime mission, we gain 40 (20) times as much solid angle for a magnitude gain of 2.67 (2.1) mag or a factor of 12 (10) in brightness. These brighter stars would be easier to follow-up with spectroscopy and high resolution imaging for confirmation and subsequent characterization. Operation of the Kepler 2-wheel mode starting in 2014 would allow follow-up and characterization observations with Spitzer and HST in the optical and near-IR.

Among the most compelling targets for future exoplanet spectroscopy are planets in the 1.5-2.5 $R_\oplus$ (super Earths to Neptunes) orbiting in the habitable zones (HZ) of their host stars. Late K and M stars are valuable host stars as their smaller stellar radii and shorter HZ periods make small HZ planets easier to detect. To date Kepler has found a few cool star ($T_{eff}$<4000 K) planetary systems, with 12 confirmed systems and 60 candidates in the NASA Exoplanet Archive. Of these 72 systems, Kepler in its 2-wheel mode would recover 63 of these (87%) even with the new limiting precision (single transit SNR$\geq$3 and SNR$\geq$7 with 3 or more transits in a 30 day period) because the transit depths are large with an average of 2500 ppm. With a survey strategy optimized to observed 40 (20) new fields for 30 (60) days each over the course of ~3 years, then we would have as many as 500 new M star planets (Table 2). While these stars are relatively faint at visible wavelengths, they are quite bright in the 3 - 5 μm region that JWST will probe with [4.5] = 3-4 mag brighter than the R magnitudes cited in Figure 4.

**Table 2. Results of Survey of 40 Kepler Fields**

| Spec Type | # Stars | Avg. R Mag | Avg. Dist (pc) | # Gas Giant Planet | Avg. Mass (M) | Avg. Orbit (AU) | # Rocky Planet | Avg. Mass M$_\oplus$ | Avg. Orbit (AU) | Avg. Tplanet (K) |
|---|---|---|---|---|---|---|---|---|---|---|
| A3 | 1,200 | 8.2 | 170 | 1 | 0.124 | 0.12 | 0 | 0.0 | 0.00 | 1616 |
| F5 | 7,500 | 9.1 | 227 | 4 | 0.327 | 0.04 | 0 | 0.0 | 0.00 | 1938 |
| G0 | 16,000 | 9.7 | 227 | 15 | 0.051 | 0.03 | 0 | 0.0 | 0.00 | 1794 |
| G5 | 24,000 | 11.0 | 213 | 28 | 0.073 | 0.04 | 0 | 0.0 | 0.00 | 1428 |
| K0 | 28,000 | 11.4 | 208 | 42 | 0.042 | 0.04 | 1 | 4.5 | 0.01 | 1290 |
| K3 | 24,000 | 12.6 | 236 | 46 | 0.046 | 0.04 | 2 | 8.1 | 0.01 | 1145 |
| K5 | 29,000 | 13.0 | 223 | 52 | 0.07 | 0.04 | 5 | 7.6 | 0.01 | 1057 |
| M0 | 12,000 | 11.7 | 155 | 35 | 0.023 | 0.04 | 8 | 7.0 | 0.02 | 787 |
| M2 | 28,000 | 14.0 | 179 | 67 | 0.035 | 0.04 | 48 | 6.3 | 0.03 | 657 |
| M3 | 20,000 | 14.0 | 133 | 38 | 0.026 | 0.03 | 60 | 5.0 | 0.03 | 602 |
| M4 | 81,000 | 15.5 | 157 | 132 | 0.038 | 0.03 | 230 | 4.4 | 0.03 | 500 |
| M5 | 33,000 | 16.5 | 130 | 43 | 0.035 | 0.03 | 87 | 4.4 | 0.03 | 380 |
| M6 | 37,000 | 17.7 | 130 | 44 | 0.030 | 0.03 | 67 | 4.5 | 0.03 | 335 |
| M7 | 2,200 | 17.3 | 64 | 2 | 0.023 | 0.02 | 4 | 3.9 | 0.02 | 343 |
| M8 | 2,600 | 18.9 | 68 | 4 | 0.022 | 0.02 | 4 | 4.6 | 0.02 | 339 |
| Tot. | 346,500 | | | 548 | | | 511 | | | |

We can estimate the yield of such a survey by incorporating: 1) the assumed characteristics of Kepler in its extended mission as described above and instrumental noise (Jenkins et al (2010); 2) a population of ~4,000 bright FGK star with R<14 mag in each 100 sq. deg. field; 3) a population of ~4,000 M stars ranging in brightness from R<14 mag for M0 stars down to R<18 mag for M8 stars in each 100 sq. deg. field (Figure 4); 3) planet period and radius distribution functions from Howard et al (2012) and Petigura et al (2013), including a flattening in the occurrence rate of planets smaller than 2 $R_\oplus$, a cutoff in the occurrence of gas giants larger than 0.5 $M_{Jup}$ for M stars, and an overall probability of 50 % that a star hosts at least one planet larger than 1 $R_\oplus$; 4) standard criteria for acceptance of 3 or more transits with final SNR greater than 7.

Preliminary analysis using the 2MASS catalog suggests there will be roughly 4,000 M stars in each 100 sq. deg. Kepler Field of view with Ks<13 mag (R <18 mag) of which 1,000 stars will be bright than R~ 15 mag. These are predominantly M0-M5 stars with a smattering of later spectral types. By looking in fields studied in greater depth with the Sloan Digital Sky Survey (West et al 2008), we can extend the sample with 150-250 M5-M8 stars per Kepler FOV which, although fainter than other M stars (SDSS i<18 mag) would have deep transits for even small planets.

In parallel with the M star observations, Kepler would also target bright FGK stars to fill the gap in stellar brightness between Kepler and ground based surveys. There are roughly 4,000 FGK main sequence stars brighter R~14 mag in 100 sq. deg. at a galactic latitude of 45 deg. Thus, in a series of forty 30-day surveys, we could study 150,000 such stars to find planets ranging in size between Jupiters to Neptunes. While these planets would be relatively hot, they would be ~10 times brighter than the typical Kepler star and thus excellent targets for future spectroscopy.

Table 2 and Figure 3 describe the results of a simulated survey of forty Kepler fields. There are relatively few planets found around the FGK stars compared with Kepler's prime mission because the large stellar radii and relative scarcity of Jovian sized planets make detection of planets with accessible transit depths difficult. However, when such planetary systems are found, they orbit quite bright stars , R~10 mag. Where Kepler would excel is in targeting M stars where the photon limited SNR is high enough to detect modest transit depths from small planets which Kepler's prime mission results suggest are abundant around low mass stars. Accordingly, our simulation suggests we would find hundreds of planets orbiting nearby M stars. A 30 (60) day stare at each field results in a significant number of habitable zone planets with effective temperatures of 400 (300) K.

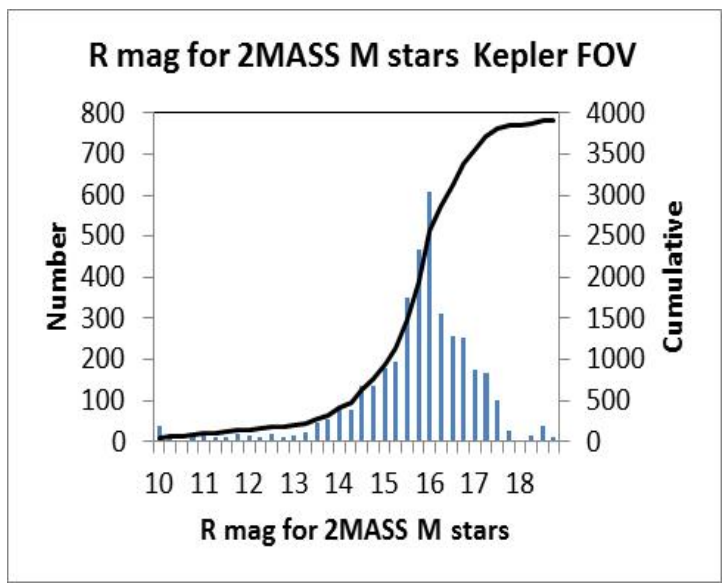

Figure 4. Predicted R magnitudes for color-selected 2MASS M stars in a 10x10 field. The bars show the break down by magnitude while the solid curve shows cumulative distribution on the right-hand scale. The turnover in the source counts reflects the 2MASS magnitude limit and is thus a conservative estimate of the number of available targets.

The combined survey of forty fields, each containing ~4,000 FGK stars (R<14 mag) and ~4,000 M stars (R<14-18 mag depending on spectral type) , and observed for 30 days would monitor in total 320,000 stars. This is roughly double the size of Kepler's prime mission target list, but for stars which are typically 10 times brighter. The yield of a simulated survey (Figure 3 and Table 2) includes over 500 gas or icy giants and approximately 500 rocky planets. The latter have masses between 2-5 $M_\oplus$ and equilibrium temperatures between 250 and 600 K.

It is important to note that the challenge of confirmation and characterization will be greatly reduced relative to the Prime mission. The stars will be 10× brighter for easier spectroscopic and imaging observations.  M star targets will have K<13 in the near-infrared and will be within reach of the upcoming  generation of precision near-infrared radial velocity instrumentation, including iSHELL on IRTF, CARMENES, the Habitable Zone Planet Finder at HET, NIRSPEC on Keck, and the cross-dispersed CRIRES on the VLT. Equally important, the background density will be a factor of 10 lower at high latitudes. For example, in a representative portion of the Kepler field, the UCAC-2 catalog lists 24,400 stars/sq.deg to R<18 mag in the middle of Kepler field compared to 2,100 stars/ sq.deg to R<18 mag at a galactic latitude of 45 deg. These source densities correspond to 0.98 background source per Kepler PSF in the nominal mission compared with 0.08 per Kepler PSF at high latitudes. These background objects are in many cases responsible for the false positive detections that must be carefully rejected with a sustained observing program.

**Survey Known Radial Velocity Systems For Transits.** Finally, during those times when seasonal constraints put the ecliptic plane closer than desired to the galactic plane (Figure 1), Kepler could target fields centered on stars hosting short period planets discovered with radial velocity (RV) observations. The NASA Exoplanet Archive lists 45 planets with M sin(i) >0.1 $M_{Jup}$ and periods <30 days discovered via RV. With advance knowledge that these planets exist and armed with well determined orbits Kepler could stare for ±1/8 of a period or less, just a few days per star, to see if the known planet is transiting. The yield of this survey would be completely confirmed and characterized plants orbiting bright stars (average V=7.5 mag with V ranging from 4.1 to 10 mag) and ready for immediate spectroscopic characterization. With a transit probability of $\eta \approx \sim (R_*/a)/(1-\varepsilon^2)$ where R is stellar radius, a is the planet semi-major axis and $\varepsilon$ the orbital eccentricity, we can estimate the number of transiting planets in a survey of previously identified RV planets. In addition, if we estimate conservatively that we need to search a given star for ±1/8 of the orbital period of its planet, then we can also calculate the cumulative number of days needed to search N bright stars. Figure 1 shows a sampling of known RV systems with a high transit probability (>5%) within ±30 degrees of the ecliptic. Figure 5 shows the cumulative number of planets and the total duration of this program. These confirmed and well characterized planets orbiting bright stars would be invaluable for spectroscopy with JWST and other facilities.

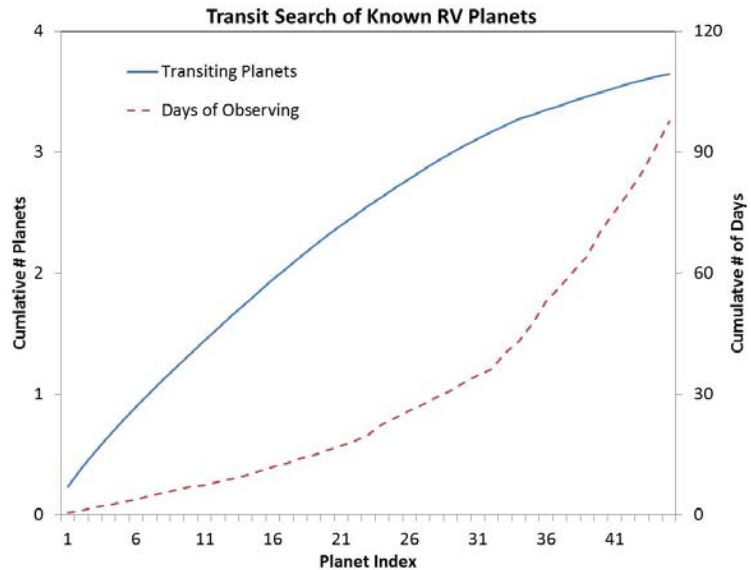

Figure 5. The cumulative number of transiting planets expected in a target survey of known RV systems (solid line). The planets are ordered in descending probability of transit. The dashed line shows that by studying ~30 stars, each for a duration of ±1/8 of its period, we could expect to detect 3 planets in just 30 day program. The expected number of planets flattens out quickly while the required number of days increases sharply as stars of lower probability (larger orbital separation and/or+ lower eccentricity) are examined.

## B. Star Formation Science

During its main mission, Kepler stared continuously at a single field of giant and main sequence stars, the youngest of which were ~380 Myr. The lack of pre-main-sequence targets was intentional, as the high levels of variability associated with young stellar objects (YSOs) presents a challenge to transit detection. However, there is important science to be gained from including star forming regions (SFRs) in a new set of Kepler fields. Continuous, high precision photometric monitoring by the *MOST* and *CoRoT* satellites has demonstrated recently that optical light curve variability in YSOs involves a surprising variety of phenomena, from fading events attributed to non-uniformities in the circumstellar disk to stochastic accretion and spots on the star and/or disk (e.g., Alencar et al., 2010, Siwak et al. 2011, Cody & Hillenbrand 2013, Cody et al. 2013). Time domain photometric studies are leading to new insights, but larger samples with longer, more continuous stare times and higher precision than is possible from the ground (Figure 6) are needed to probe the range and frequency of these phenomena.

We highlight two science goals that could be accomplished by pointing Kepler toward nearby star forming regions: 1) better understanding of YSOs and their inner disks through photometric monitoring, and 2) identification of young, low mass eclipsing binary systems to serve as benchmarks for stellar evolution models. The discovery of youthful transiting planets in these fields is a further possibility, but with only one transit candidate associated with a pre-main-sequence star (van Eyken et al. 2012; Barnes et al 2013), the probability of detection around active and accreting young stars is difficult to assess.

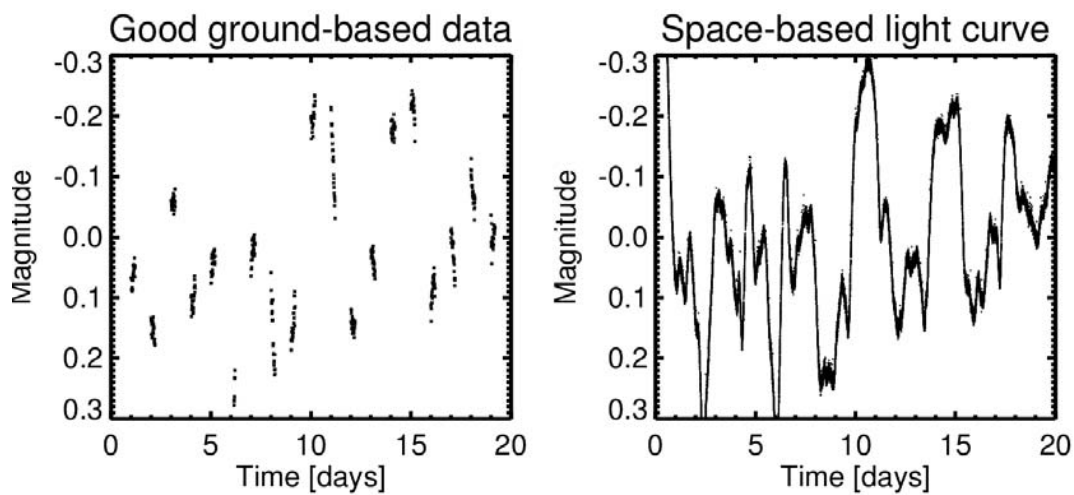

Figure 6. Comparison of space data from the *CoRoT* satellite toward a YSO (right) with a degraded version of the same data (left) as they might be seen from the ground under good conditions.

**A More Complete Characterization of Optical Variability in Young Disk-bearing Stars.** YSOs display optical variability on timescales from hours to years, at amplitudes from the percent level to over a magnitude. The photometric behavior of classical T Tauri stars is often erratic and lacking in stable periodicities. To fully mine the light curve phenomenology and connect it to (circum)stellar physics, we face three requirements: 1) nearly continuous monitoring for several stellar rotation periods (i.e., a few weeks); and 2) photometric precision at the <<1% level or better on stars down to V~16; and 3) sufficient field of view size to monitor a large number of YSOs simultaneously. Kepler is currently the only telescope that can meet all of these criteria *in the optical*.

Current and planned observing campaigns involving YSOs meet some of the above requirements, but fall short in providing a full window into their time domain behavior. The Palomar Transient Factory

(Law et al. 2009) has monitored variable young stars in areas of Orion and the North American Nebula

| Table 3. Nearby star-forming regions for Kepler monitoring | | | | | |
|---|---|---|---|---|---|
| Region | Known Members | Spatial size | Magnitude Range | Ecl. Latitude (deg) | Distance (pc) | Age (Myr) |
| Taurus | ~400 | ~200$^{2°}$ | $V$=9-17 | -3.6 | 145 | 1-3 |
| Upper Scorpius | ~400 | ~250$^{2°}$ | $V$=9-17 | -2.1 | 145 | 5-6 |
| Orion complex | >5000 | ~80$^{2°}$ | $V$=12-21 | -28 | 400-450 | 1-10 |
| Ophiuchus | ~300 | ~25$^{2°}$ | $V$=11-20 | -1.7 | 120 | 1-3 |

with its 7.8$^{2°}$ field of view, but the time sampling from this shared ground-based facility is highly non-uniform (Findeisen & Hillenbrand 2013). SFRs will also be observed with the Large Synoptic Survey Telescope campaign, but at a relatively poor cadence of once every few days. Other space telescopes are inadequate for the task of monitoring most YSOs: *CoRoT* recently ceased operation; *MOST* can only reach stars brighter than V=13; and the fields of view available through HST are too small to monitor a statistically significant number of YSOs. A comparison of a disk-bearing T Tauri star light curve from the *CoRoT* mission with resampled data as expected from a good ground-based observing run (i.e., no substantial weather interruptions) is shown in Figure 6. This simulation confirms that space-based monitoring is essential to capturing the full complexity of photometric behavior in YSOs.

We note that the Young Stellar Object Variability project (YSOVAR; Morales-Calderón et al. 2011) has made headway into classifying the *infrared* flux variations of YSOs down to the 1% photometric level, but often this behavior is uncorrelated with that seen in the optical. The Coordinated Synoptic Investigation of NGC 2264 (CSI 2264) carried out simultaneous *CoRoT* and *Spitzer* monitoring of YSOs, discovering that less than half of disk bearing objects display a simple wavelength dependence between their optical and infrared light curves (Cody et al. 2013). Clearly, these two bands probe different physics from the stellar surface and accretion shock (optical/UV) to the inner disk (near to mid-infrared). Given that NGC 2264 is the *only* young cluster with more than a handful of stars monitored continuously at high cadence and precision, observations with Kepler would greatly expand the morphological catalog of optical YSO light curve behavior.

We list in Table 3 several SFRs that Kepler could target as part of its extended campaign. The telescope is ideal for observing the more dispersed (>1°) regions that have typically been difficult to monitor from the ground. We note that while some of the SFRs occupy an area larger than the Kepler FOV, the populations tend to be clustered. In particular, the Taurus and Upper Scorpius each have a few hundred known members within a 10°x10° low extinction region (Ardila et al. 2000, Preibisch et al. 2002, Slesnick et al. 2006), and they are on the ecliptic where pointing (and hence photometry) is predicted to be most stable. A further advantage of targeting these nearby regions is that the magnitude range of T Tauri stars (primarily $V$~11-15) makes them ideal for follow-up with many types of ground-based facilities. We envision using Kepler to conduct a time domain survey of YSOs, after which the community could pursue imaging or spectroscopic observations on the objects with the most interesting light curves.

**Search for Young Eclipsing Binary Systems.** There are only eight published young eclipsing binary (EB) systems in which both components are in the coveted low mass regime (< 1.5 $M_\odot$); the majority are in Orion and in the 1-2 Myr range (e.g. Morales-Calderón et al. 2012). In addition, two short runs with the *CoRoT* satellite have produced a handful of new candidate EBs in the NGC 2264 cluster (Gillen et al. 2013, submitted). Eclipsing binaries are crucial for testing and calibrating theoretical stellar models in the 1-10 Myr range, in which evolution proceeds rapidly. Comparison of low-mass EB parameters with theoretical radius and mass predictions has revealed differences of order ~20%. Additional systems, particularly in the 3-10 Myr regime, are needed to assess the reasons behind this discrepancy, which could include enhanced magnetic activity and suppressed convection. A monitoring survey of several hundred SFR members is likely to pick up at least a few young, low-mass EBs, since eclipses depths of 10% or more would easily captured by Kepler, even in the worst case photometric performance. These data are of fundamental importance in calibrating the models presently being used to infer the properties of young, planetary mass objects now being discovered in direct imaging observations.

## IV. Implementation

By necessity we will need to balance the number of targets observed, the number of pixels dedicated to each target, and the frequency with which we download the accumulated data. The mission currently observes ~3.8 million pixels per 29.4-minute cadence, and downloads the data every ~30 days. In the best-case pointing stability scenario presented by the project, along the velocity and anti-velocity vectors of the spacecraft in the ecliptic plane, the boresight drift can be limited to <1.25 pixels (5 arcseconds) per day with daily momentum management. On its own, this drift would not significantly increase the typical number of pixels required per target – for instance, a 12th magnitude target typically contains ~20 pixels in the optimal aperture. With foreknowledge of the direction of the drift this would only need to be increased by 25%. However, with the pointing precision limited by the star trackers to 15-30 arcseconds, the size of the optimal aperture would need to be more than doubled (~40-50 pixels) in order to ensure the flux from the target was captured, implying a reduction in the number of targets able to be observed by a factor of ~2.5 to ~60,000, if the average magnitude of the observed targets remained the same. Since observing brighter targets is an ideal way to recover Kepler's photometric sensitivity, this would involve a further reduction in targets by another factor of ~2.5 to increase the average brightness by 2 magnitudes, to ~24,000.

Transits of exoplanets around M-dwarf targets have relatively short ingress and egress times on the order of 10 minutes. Although constraining these times is not essential for discovery (c.f., the discovery of the Kepler-32 system), they are useful for the characterization of the exoplanet properties. Thus we propose to decrease the Kepler cadence time from 29.4 to 14.7 minutes which would result in approximately another factor of two decrease in the number of targets that could be observed to approximately 12,000. *This number is adequate to observe the ~8,000 bright FGK and M stars per field as described above, including over-selecting stars to be able to reject giant stars based on in-orbit data. It will also be possible to add an additional ~4,000 targets as provided by the community.*

The numbers in the previous section are based on the assumption that the apertures defined for each target are constant over the observing period, and large enough to account for both the imprecision in the pointing due to the star trackers and the pointing drift. Another option is that, with foreknowledge of the direction of the drift, the aperture be allowed to 'drift' over the observing period along with the targets. This would require substantial changes to the flight software and ground-based processing software, but would allow us to recover a larger number of targets, perhaps 2-3 times as many. However, given that the science outlined can be achieved with several thousand targets per field of view, we suggest it may be preferable to define larger apertures and maintain the current software.

We anticipate that the earliest steps in the Kepler pipeline, e.g. the CAL module for pixel calibration using pre-launch flat-field data and parts of the PA module for cosmic ray removal and simple aperture photometry can be used for this new operational mode. These routines can be used with relatively little modification to extract pixel data from the downlinked data stream, adding appropriate header information, and carrying out basic instrument calibrations. The resultant postage stamp data would be placed into the MAST archive for use by the community using standard CCD photometry algorithms to achieve ~1 mmag photometry. This goal, while challenging, is certainly possible with space based data as demonstrated with the EPOXI data (Ballard et al 2010). Subsequent data reduction and analysis would be performed by the individual investigators.